\documentclass[aps,prb,amsmath,amssymb,showpacs,superscriptaddress,12pt]{revtex4-1}

\usepackage{graphicx}
\usepackage{amsmath,amssymb,amsfonts}
\usepackage{textcomp}
\usepackage{hyperref}
\usepackage{gensymb}
\usepackage{verbatim}
\usepackage{amsmath}
\hypersetup{breaklinks=true,colorlinks=true,urlcolor=black}
\usepackage{color}
\usepackage{soul}
\DeclareGraphicsExtensions{.jpg,.pdf,.png,.eps}

\begin{document}
\title{Intrinsic axion insulating behavior in antiferromagnetic MnBi$_6$Te$_{10}$}

\author{Na Hyun Jo}
\email[]{njo@iastate.edu}
\affiliation{Ames Laboratory, Iowa State University, Ames, Iowa 50011, USA}
\affiliation{Department of Physics and Astronomy, Iowa State University, Ames, Iowa 50011, USA}

\author{Lin-Lin Wang}
\email[]{llw@ameslab.gov}
\affiliation{Ames Laboratory, Iowa State University, Ames, Iowa 50011, USA}

\author{Robert-Jan Slager}
\email[]{rjslager@g.harvard.edu}
\affiliation{Department of Physics, Harvard University, Cambridge, Massachusetts 02138, USA}

\author{Jiaqiang Yan}
\affiliation{Materials Science and Technology Division, Oak Ridge National Laboratory, Oak Ridge, Tennessee 37831, USA}

\author{Yun Wu}
\affiliation{Ames Laboratory, Iowa State University, Ames, Iowa 50011, USA}
\affiliation{Department of Physics and Astronomy, Iowa State University, Ames, Iowa 50011, USA}

\author{Kyungchan Lee}
\affiliation{Ames Laboratory, Iowa State University, Ames, Iowa 50011, USA}
\affiliation{Department of Physics and Astronomy, Iowa State University, Ames, Iowa 50011, USA}

\author{Benjamin Schrunk}
\affiliation{Ames Laboratory, Iowa State University, Ames, Iowa 50011, USA}
\affiliation{Department of Physics and Astronomy, Iowa State University, Ames, Iowa 50011, USA}

\author{Ashvin Vishwanath}
\email[]{avishwanath@g.harvard.edu}
\affiliation{Department of Physics, Harvard University, Cambridge, Massachusetts 02138, USA}

\author{Adam Kaminski}
\email[]{kaminski@ameslab.gov}
\affiliation{Ames Laboratory, Iowa State University, Ames, Iowa 50011, USA}
\affiliation{Department of Physics and Astronomy, Iowa State University, Ames, Iowa 50011, USA}

\date{\today}
\maketitle 
{\bf
A striking feature of time reversal symmetry (TRS) protected topological insulators (TIs) is that they are characterized by a half integer quantum Hall effect on the boundary when the surface states are gapped by time reversal breaking  perturbations. While time reversal symmetry (TRS) protected TIs have become increasingly under control, magnetic analogs are still largely unexplored territories with novel rich structures. In particular, topological magnetic insulators can also host a quantized axion term in the presence of lattice symmetries. Since these symmetries are naturally broken on the boundary, the surface states can develop a gap without external manipulation. In this work, we combine theoretical analysis, density functional calculations and experimental evidence to reveal intrinsic axion insulating behavior in MnBi$_6$Te$_{10}$. The quantized axion term arises from the simplest possible mechanism in the antiferromagnetic regime where it is protected by inversion symmetry and a fractional translation symmetry. The anticipated gapping of the Dirac surface state at the edge is subsequently experimentally established using Angle Resolved Spectroscopy.  As a result, this system provides the magnetic analogue of the simplest TRS protected TI with a single, gapped Dirac cone at the surface. 
}

Topology has drastically changed our understanding of insulators and metals. One of the most intriguing consequences of topological materials is that they allow for the realization of exotic quasiparticles and phases in real materials in the laboratory. A prime example in this regard is that TIs provide for the condensed matter realization of the $\theta$-vacuum\cite{Qi_2008,Wilczek1987}, meaning that the electromagnetic response is governed by a term
\begin{equation}
 S_{\rm axion} = \theta \frac{\alpha}{(4\pi^2)} \int d^3xdt \,\mathbf{E}\cdot\mathbf{B}, 
\end{equation}
where $\alpha=e^2/(\hbar c)$ refers to the fine structure constant. This term is manifested through the topological magnetoelectric effect (TME), having profound consequences\cite{Wilczek1987, Essin2009,Qi_2008}. In particular, the response coefficient is quantized to odd numbers of the fine structure constant $\alpha=e^2/(\hbar c)$ divided by $4\pi$. Being a total derivative, the TME finds its physical origin by considering the edge. Here, its presence results in a half quantized anomalous Hall conductivity, which essentially corresponds to the odd number of Dirac fermions on the surface \cite{Qi_2008}. The exotic nature of the TME also has been associated with other proposals, such as inducing a magnetic monopole upon bringing electric charge close to the surface \cite{Qi2009}, which in some scenarios is proposed to allow for the formation of the condensed matter variant of the elusive dyon excitation \cite{Witten1979} as well as magneto-optical Faraday and Kerr rotation effects \cite{Chang2009,Maciejko2010,Tse2010,Wu1124}.

While upon gapping the surface states of 3D TRS protected TIs, having $\theta=\pi$, the TME can be accessed, this requires intricate external manipulation.
On the other hand, magnetism cannot only induce novel topological structures \cite{PhysRevB.83.205101} that are yet to be fully explored, but may give rise to axion and quantum anomalous Hall effects {\it intrinsically} \cite{Wan2012,PhysRevB.88.121106}. In particular, the presence of lattice symmetries, such as fractional translations combined with time reversal symmetry, or inversion symmetry, can also give rise to a quantized axion angle, $\theta=\pi$. As surface terminations will generally break these symmetries, the associated gapless states of such magnetic TIs are naturally gapped without the need for complicated external manipulation. As a result such systems provide a natural platform for the aforementioned physical effects. Finally, magnetic domain walls of topological magnetic insulators can host conducting modes making them an exciting platform to study in their own right \cite{Ma2015}.


Recently, it was predicted that this physics can be realized in MnBi$_{2}$Te$_{4}$ \cite{Zhang2019}. While some ARPES studies reported observation of a gapped surface state and thus possible axion insulator behavior\,\cite{zeugner2019, Lee2019, Vidal2019}, many others showed evidence that the surface state in this material remains gapless down to low temperatures \,\cite{swatek2019gapless, chen2019,hao2019,chen2019Topological,li2019dirac}. The latter may be a result of weak hybridization between the magnetic states and the
topological electronic states\,\cite{li2019dirac} or possibly a different (or disordered) magnetic structure at the top Mn-Te layer\,\cite{swatek2019gapless, hao2019}.

In this work, we report theoretical understanding and experimental evidence of intrinsic axion insulating behavior in a different but related material, MnBi$_6$Te$_{10}$ with A-type AFM order, having a reported bulk transition temperature of $T$\,=\,11\,K.\,\cite{yan2019}. MnBi$_{6}$Te$_{10}$ has a rhombohedral crystal structure (R$\bar{3}$m, 166) with repetitions of a septuple layer and two quintuple layers. 
The topological evaluation of the band structure is remarkably transparent, see also SI \ref{app:classifcation}. Due to the inversion symmetry $I$ of R$\bar{3}$m the properties are conveniently determined by analysis of the parity eigenvalues at the time reversal invariant momenta (TRIM) satisfying $\mathbf{k}=n_{1}/2\mathbf{b}_1+n_{2}/2\mathbf{b}_2+n_{3}/2\mathbf{b}_3$, where $\mathbf{b}_i$ comprise the reciprocal lattice vectors and $n_{i}$ are integers. Generalizations of the Fu-Kane criterion \cite{Fu_2007} have not only recently proven to be powerful in the context of evaluating topological band structures \cite{Po_2017, Watanabe_2018, Slager_2013, Kruthoff_2017, Bradlyn_2017}, but can also reveal the presence of a theta term as well as Chern numbers \cite{Turner_2012, Hughes_2011}.

In the paramagnetic phase, application of  the Fu-Kane criterion \cite{Fu_2007} reveals that the system is a time reversal symmetry (TRS)  protected strong topological insulator (TI).  Physically, we find that the parity even/odd $|P_z ^+ (\uparrow\downarrow) \rangle$ and $|P_z ^+ (\uparrow\downarrow) \rangle$ states invert at the $\Gamma$ point due to spin-orbit coupling, leading to the topological insulator phase. This is similar to TRS
protected TIs, such as $\text{Bi}_2\text{Se}_3$ and $\text{Bi}_2\text{Te}_3$, that exhibit a single cone at the surface \cite{Zhang_2009} due to a band inversion at the $\Gamma$ point, giving one odd pair in the parity evaluation.

In the AFM phase the unit cell doubles, and TRS, represented by $\Theta$, is broken, removing the necessity of Kramers pairs. Nonetheless, a symmetry $S$ is generated that is the product of a fractional translation, $T_{1/2}=1/2(\mathbf{b}_1+\mathbf{b}_2+\mathbf{b}_3)$, and TRS, $S=\Theta T_{1/2}$. $S$ squares to a lattice translation and thus depends on momentum \cite{Mong_2010}. Defining $\tilde{k}=T_{1/2}\cdot\mathbf{k}$, we then observe that $S^2=-1$ for $\tilde{k}=0$ and  $S^2=1$ for $\tilde{k}=\pi$. Consequently, the $\tilde{k}=0$ plane is characterized by a $\mathbf{Z}_{2}$ index as $S$ acts as pseudo-TRS\cite{Mong_2010}. This fits within a generalized perspective on $I$-symmetric insulators, in which one evaluates half of the difference between the number of filled parity even $n^+$ and parity odd $n^-$states at the TRIM \cite{Turner_2012},  
\begin{equation}\label{eq::indicator}
\eta=1/2\sum_{\mathbf{k}\in \Gamma_i} \left ( n_\mathbf{k}^+-n_\mathbf{k}^- \right ) \qquad \text{Mod }4.
\end{equation}
This quantity is related to a $\mathbf{Z}_4$ index as a trivial bands can be added. Whereas the odd outcomes $\eta=1,3$ diagnose Weyl semimetals, $\eta=0,2$ can correspond to insulating phases. The difference between a trivial ($\theta=0$) and axion  ($\theta=\pi$) insulator is that the latter  requires a distribution of parity eigenvalues that amounts to twice an odd integer number $\eta$ \cite{Turner_2012}. Potentially, this may also correspond to a stacked Chern insulator. As explained in SI \ref{app:classifcation}, the presence of $S$ symmetry however ensures that parity eigenvalues comes as pairs at TRIM, eliminating the stacked Chern insulator possibility and reducing the characterization to the anticipated $\mathbf{Z}_2$ index.
Due to the single inverted pair at the $\Gamma$-point, we conclude that the system is a  quantized axion  ($\theta=\pi$) insulator and may be regarded as the AFM analog of the simplest TRS topological insulator with a single cone at the surface \cite{Zhang_2009}.

$S$ symmetry thus rather acts as TRS, although being subtly different. As in the TRS case, the distribution of parity eigenvalues allows for determination of the $\mathbf{Z}_2$ index and edge states can be gapped when $S$ symmetry is broken \cite{Varnava_2018}. In contrast to TRS this however depends on the surface termination. Accordingly, we show that natural cleavage planes in the (001) direction a gap develops, that is associated with magnetic ordering. In contrast $S$ preserving surfaces show persistent gapless edge states, see SI \ref{app:DFT}. We note that ferromagnets in presence of $I$ symmetry can similarly exhibit a axion phase, but the possibility of a surface preserving $S$ symmetry makes the edge theory more versatile from the viewpoint of TME phenomena, as any termination breaks $I$ symmetry.

\begin{figure}[!ht]
	\includegraphics[width=5 in]{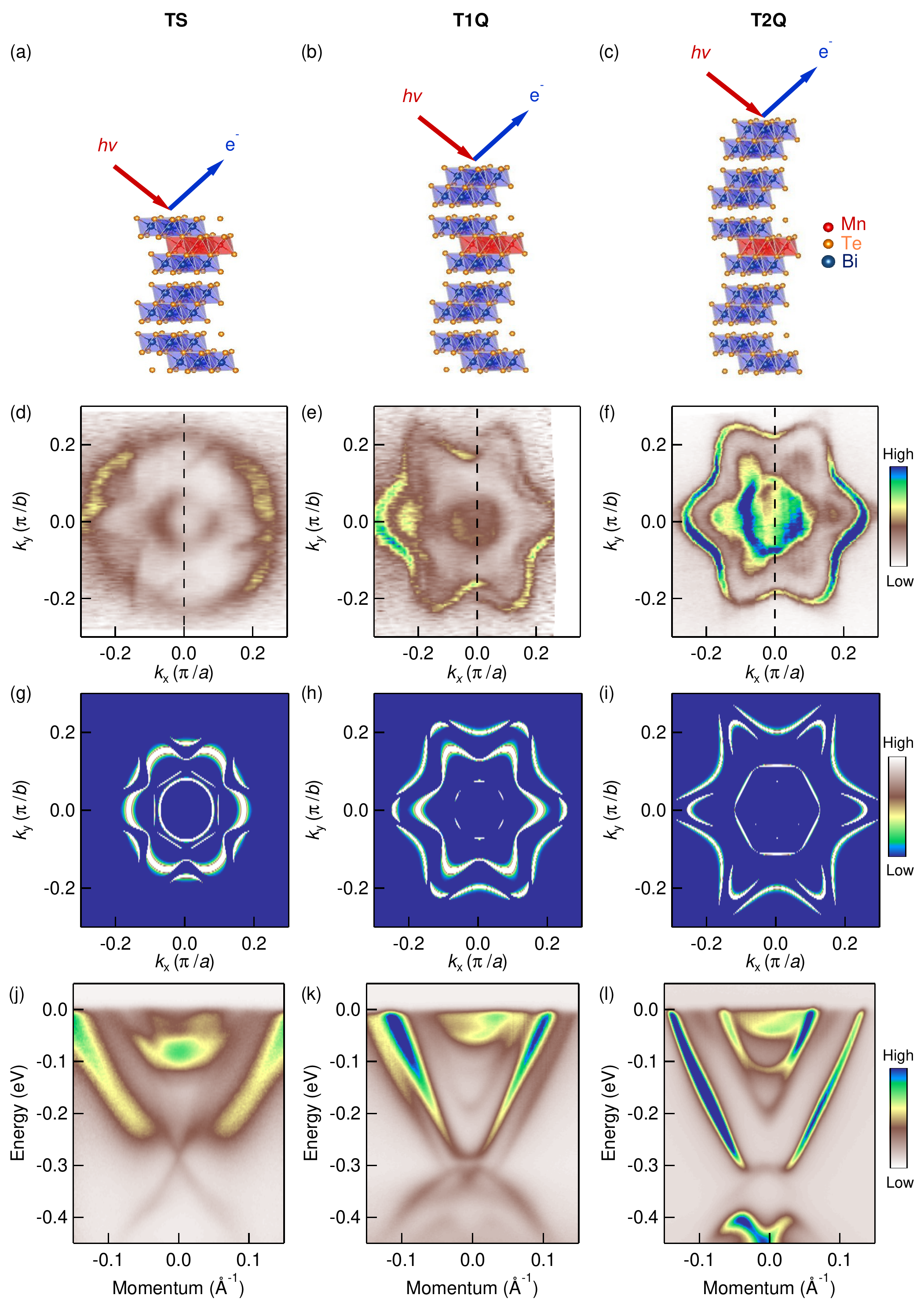}%
	\caption{\textbf{Crystal structure and band structures of MnBi$_{6}$Te$_{10}$ with three different terminations} \textbf{a-c}, Crystal structure of MnBi$_{6}$Te$_{10}$ and possible terminations, TS, T1Q and T2Q. \textbf{d-f}, Fermi surface plots of the TS, T1Q, and T2Q terminations, respectively. \textbf{g-i}, DFT calculations of Fermi surfaces of the TS, T1Q, and T2Q terminations, respectively. \textbf{j-l}, Band dispersions along the high symmetry lines of $k_{x}$\,=\,0 of (b)-(d).}
	\label{fig:1}
\end{figure}


 Due to the weak van der Waals bonding between Te-Te layers, MnBi$_{6}$Te$_{10}$ can exfoliate above a septuple layer (TS termination, Fig.\,\ref{fig:1} (a)), a qiuntuple layer (T1Q termination, Fig.\,\ref{fig:1} (b)), and two qiuntuple layers (T2Q termination, Fig.\,\ref{fig:1} (c)). Figures\,\ref{fig:1} (d)-(f) show three different Fermi surfaces that we observed from ARPES experiments at $T$\,=\.40\,K. We assigned the terminations based on comparison with the DFT calculations as shown in Figs.\,\ref{fig:1} (g)-(i). The determination of T1Q and T2Q terminations is clearer as we compare the energy dispersion which will be discussed later. The band dispersions along the high symmetry lines of $k_{x}$\,=\,0 at $T$\,=\.40\,K corresponding to Figs.\,\ref{fig:1} (d)-(f) are shown in Figs.\,\ref{fig:1} (j)-(l), respectively. These clearly delineate the differences between each termination. For the three different terminations, non-magnetic DFT surface band structures all give surface Dirac points (SDP). But unlike the normal situation of TS having the SDP inside the bulk band gap between band N and N+2 (N is the number of bulk valence bands), the SDPs for T1Q and T2Q are pushed downward to be between bulk band N-2 and N. At the same time, the surface conduction bands of T1Q and T2Q merge with bulk valence bands. Similar features have also been found and discussed for PbBi$_{6}$Te$_{10}$\,\cite{papagno2016}. Furthermore, with AFM ordering in DFT surface band structure calculations, there are gap openings for the SDPs on all three terminations as expected for breaking S symmetry on the (001) surface, but the size of gap is different. It is very small $\sim$\,1\,meV on T1Q and T2Q comparing to the $\sim$\,60\,meV on TS. Thus, we will focus on the gap opening of the SDP on TS below for experimentally evaluating axion behavior.  

\begin{figure}[!ht]
	\includegraphics[width=6.5in]{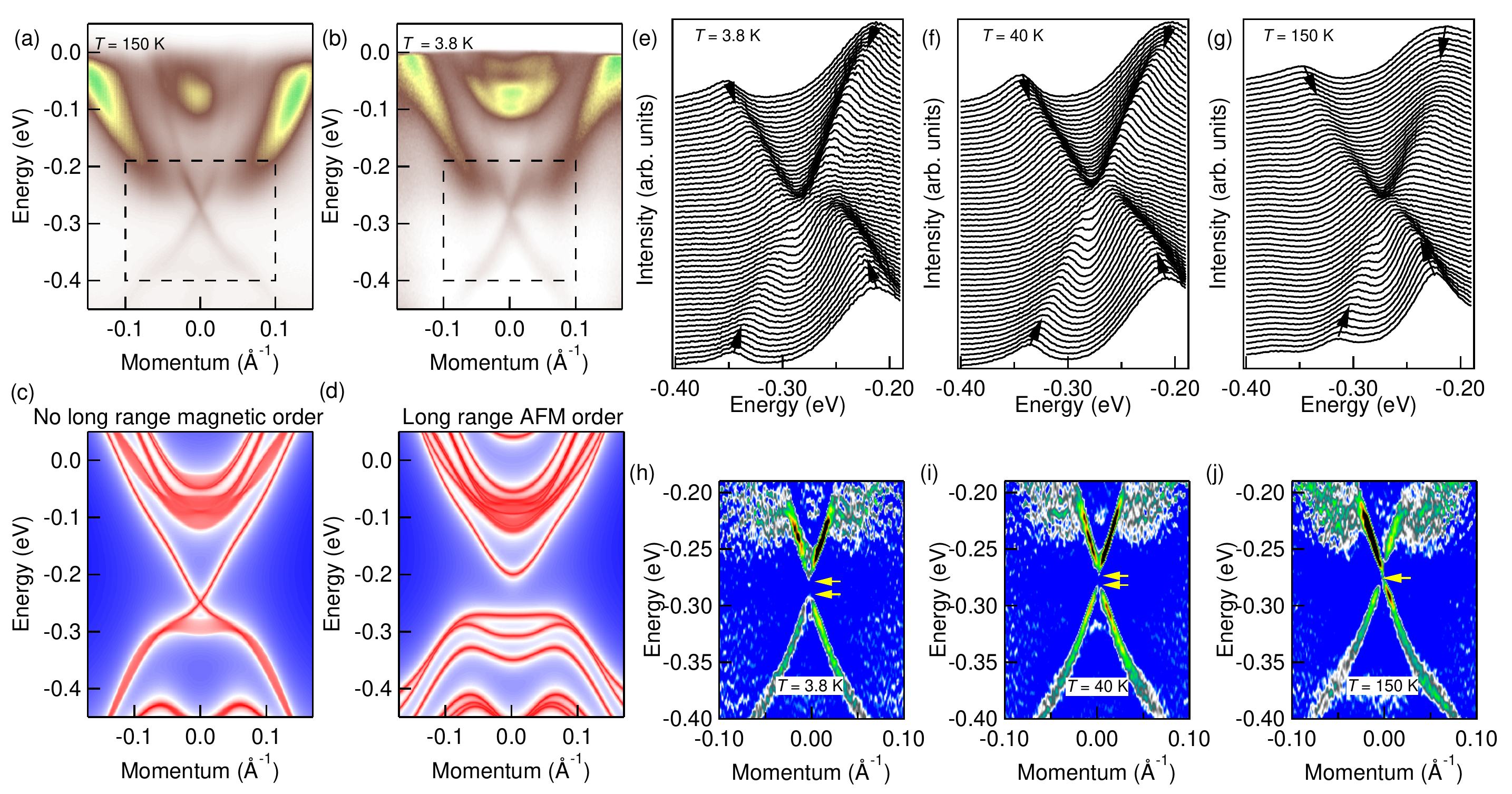}%
    \caption{\textbf{Band dispersion at the TS termination} \textbf{a-b}, Band dispersions at the TS termination at $T$\,=\,150\,K and $T$\,=\,3.8\,K. \textbf{c-d}, DFT calculations of energy dispersion with no long range magnetic order and long range A-type AFM order, respectively. \textbf{e-g}, EDCs corresponding to 1/2 of the range centered $k_{y}$\,=\,0 of the dashed box in a and b at 3.8, 40, and 150\,K respectively. \textbf{h-j}, The second derivatives of the dashed box in (a) and (b) with respect to EDC at 3.8, 40, and 150\,K.}
	\label{fig:2}
\end{figure}


Figure\,\ref{fig:2} presents the temperature dependence studies of the TS termination. Both ARPES data and DFT calculations show the topologically non-trivial surface states which marked as black dashed box in Figs.\,\ref{fig:1} (a) and (b). Although the ARPES data above and below the transition temperature (Figs.\,\ref{fig:2} (a) and (b)) seem not too much different at the first glance, the DFT calculations suggest a clear gap forming in the ordered state (A-type AFM) with a surface gap size of $\sim$\,60\,meV, which corroborates axion insulating behavior. In order to confirm the gap, we performed detailed spectroscopic experiments of the dashed box regime as indicated in the Figs.\,\ref{fig:2} (a) and (b). The energy distribution curves (EDC) show a gap at $T$\,=\,3.8\,K which is below the magnetic transition temperature as shown in Fig.\,\ref{fig:2} (e). The size of the surface gap is $\sim$\,30\,meV which is comparable to the DFT calculations. The gap still remains open at $T$\,=\,40\,K which is above the transition temperature, although the gap size reduced compared to $T$\,=\,3.8\,K. (Fig.\,\ref{fig:2} (f)) This gap finally closes at $T$\,=\,150\,K as shown in Fig.\,\ref{fig:2} (g). Opening and closing of the gap can also be observed in the second derivatives in Figs.\,\ref{fig:2} (h)-(j). The discrepancy between the transition temperature and the actual temperature that the gap closes may be a result of surface magnetic fluctuation\,\cite{Daixiang2016}. Meaning, that the magnetic reorientation is slower than electron scattering even in absence of the yet to be formed long range magnetic order. The actual transition temperature might also be slightly different, although the magnitude of the difference seems to suggest this is not the defining reason.

\begin{figure}[!ht]
	\includegraphics[width=6 in]{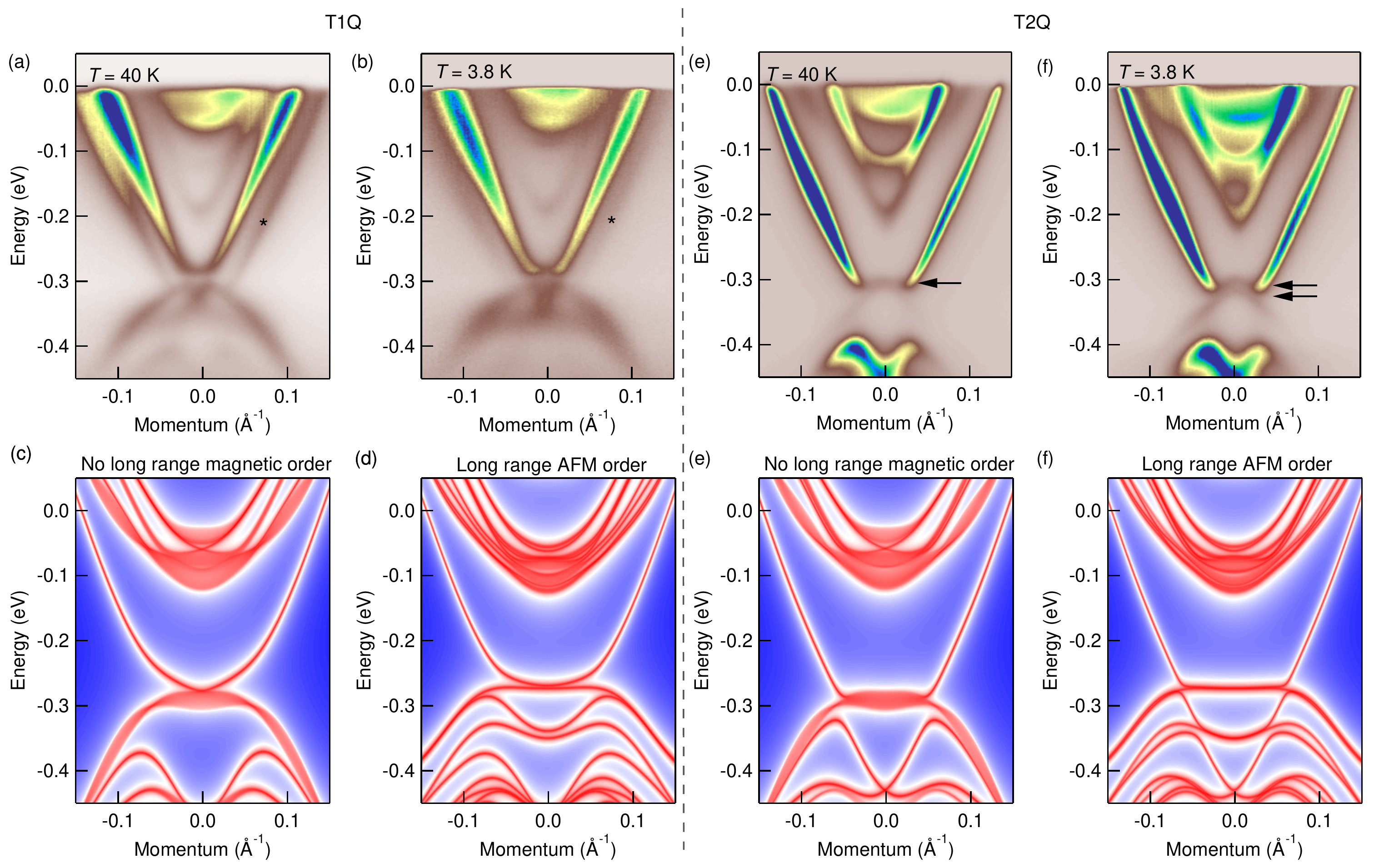}%
	\caption{\textbf{Band dispersion of the T1Q and T2Q terminations} \textbf{a-b}, Band dispersions of the T1Q termination at 40, and 3.8\,K. \textbf{c-d}, DFT calculations of the T1Q termination with no long range magnetic order and long range A-type AFM order. \textbf{e-f}, Energy dispersions of the T2Q termination at 40, and 3.8\,K. \textbf{g-h}, DFT calculations of the T2Q termination with no long range magnetic order and long range A-type AFM order.}
	\label{fig:3}
\end{figure}


Temperature dependence of band structure at T1Q and T2Q terminations are shown in Fig.\,\ref{fig:3}. For the T1Q termination, a band hybridization is detected in both above and below the transition temperature around 0.3\,eV below the Fermi energy. Note that the outer weak intensity band in T1Q termination, marked as * in Figs.\,\ref{fig:3} (a) and (b), may be the superposition of T2Q termination. Noticeable difference of energy dispersion between T1Q and T2Q terminations are the width of the strong linear surface band around 0.3\,eV below the Fermi energy. The T2Q termination has larger width than the T1Q termination. In addition, the T2Q termination demonstrates an interesting band splitting below the transition temperature as indicated by black arrows in Fig.\,\ref{fig:3} (e) and (f).  


In conclusion, we have demonstrated that MnBi$_{6}$Te$_{10}$ is one of the simplest intrinsic magnetic topological materials, making it a very promising platform to explore topological magnetic phenomena. Our studies demonstrate TRS protecting TI behavior at high temperature and intrinsic AFM axion insulating behavior at low temperatures. Future work on tuning the Fermi level at the gap by chemical doping  MnBi$_{6}$Te$_{10}$ should provide a pathway to study exciting TME effects, including a half-quantized surface anomalous Hall conductivity. 

\begin{acknowledgements}
The authors thank A. Kreyssig, C. Matt, and P. P. Orth for helpful discussion. This work was supported by the Center for Advancement of Topological Semimetals, an Energy Frontier Research Center funded by the U.S. Department of Energy Office of Science, Office of Basic Energy Sciences, through the Ames Laboratory under its Contract No. DE-AC02-07CH11358. Ames Laboratory is operated for the U.S. Department of Energy by the Iowa State University under Contract No. DE-AC02-07CH11358. Work at ORNL was supported by the U.S. Department of Energy, Office of Science, Basic Energy Sciences, Division of Materials Sciences and Engineering. 
\end{acknowledgements}

\clearpage

\section*{References}
\bibliographystyle{naturemag}

\clearpage

\section{Supplementary information}
\subsection{Methods}

Single crystals of MnBi$_{6}$Te$_{10}$ were grown using Bi-Te as flux following the procedure described in Ref.\,\onlinecite{yan2019}. Samples used for ARPES measurements were cleaved in situ at 40\,K or 3.8\,K in ultrahigh vacuum (UHV). The data were acquired using a tunable VUV laser ARPES system, that consists of a Scienta Omicron DA30 electron analyzer, a picosecond Ti:Sapphire oscillator and fourth harmonic generator.\,\cite{Rui2014} Data were collected using photon energy of 6.7\,eV. Momentum and energy resolutions were set at $\sim$\,0.005\,$\textrm{\AA}^{-1}$ and 2\,meV. The diameter of the photon beam on the sample was $\sim$\,30\,$\mu$m.

Band structures with spin-orbit coupling (SOC) in density functional theory (DFT)\,\cite{Hohenberg1964,Kohn1965} have been calculated with van der Waals exchange-correlation functional DF1-optB86b\,\cite{Dion2004, klime2011}, a plane-wave basis set and projected augmented wave method\,\cite{Bloechl1994} as implemented in VASP\,\cite{Kresse1996,Kresse1996a}. To account for the half-filled strongly localized Mn 3$d$ orbitals, a Hubbard-like U\,\cite{Dudarev1998} value of 3.0\,eV is used. For bulk band structure of A-type anti-ferromagnetic (AFMA) MnBi$_{6}$Te$_{10}$, the rhombohedral unit cell is doubled along the $c$ direction with a Monkhorst-Pack\,\cite{Monkhorst1976} (9\,$\times$\,9\,$\times$\,3) $k$-point mesh including the $\Gamma$ point and a kinetic energy cutoff of 270\,eV. Experimental lattice constants\,\cite{aliev2019} have been used with atomic positions relaxed until the absolute value of force on each atom is less than 0.02\,eV/$AA$. A tight-binding model based on maximally localized Wannier functions\,\cite{Marzari1997,Souza2001,Marzari2012} was constructed to reproduce closely the bulk band structure including SOC in the range of $E_{F}\,\pm\,1$\,eV with Mn $sd$, Bi $p$ and Te $p$ orbitals. Then the spectral functions and Fermi surface of a semi-infinite MnBi$_{6}$Te$_{10}$ (001) surface with different terminations were calculated with the surface Green’s function methods\,\cite{Lee1981, LeeJoannopoulos1981, Sancho1984, Sancho1985} as implemented in WannierTools\,\cite{Wu2018}.

\subsection{Classification Details}\label{app:classifcation}
We here discuss a few more details on the evaluation of the band structure. As noticed by early work on magnetic TIs \cite{Turner_2012}, inversion $I$ eigenvalues reduce the problem of characterizing the band structure to simply evaluating the parity eigenvalues at the TRIM. In particular if the total number of odd (-1) parties at the TRIM amounts to an odd number the systems cannot be insulating. Similarly, if the system is insulating and has vanishing Chern numbers  the parity distribution conveys whether a $\theta$ term, with $\theta=\pi$ due to $I$ symmetry, can be defined. The value of $\pi$ is then attained if the total number of odd parity eigenvalue states add to to twice an odd number. These evaluations are in essence generalizations of the Fu-Kane criterion \cite{Fu_2007}, which states that the $\mathbf{Z}_2$ invariant $\nu$  is given by
\begin{equation}
(-1)^\nu=\prod_i\delta_i  \qquad \delta_i=\prod_m \xi_{2m} (\Gamma_i),
\end{equation}
where one evaluates the parity eigenvalues $\xi$ of one of the {\it Kramer's pairs} of the $2m$ filled bands over the TRIM $\Gamma_i$, rather than individual bands as in the following paragraph.
Recently, a lot of progress has been made by considering irreps at high symmetry points in a very similar manner \cite{Slager_2013, Kruthoff_2017,Slager2019}, as these gluing condition map out a space of all possible band structures that can then be compared to Fourier Transforms of atomic band structures that by definition span the space of trivial insulators  \cite{Po_2017, Watanabe_2018, Bradlyn_2017}. When regarded as vector spaces, dividing out the trivial subspace then results in quantities that can diagnose topology arising from the underlying space group symmetry \cite{Po_2017, Watanabe_2018, Bradlyn_2017}. 

As motivated in the main text, the outcomes of parity distribution  can be diagnosed in the general light  of these symmetry indicator developments by the quantity $\eta$, 
\begin{equation}\label{eq::indicatorapp}
\eta=1/2\sum_{\Gamma_i}n_\mathbf{k}^+-n_\mathbf{k}^- \qquad \text{Mod }4.
\end{equation}
Indeed, the odd values identify Weyl semi-metals, as these parity configurations show that Billouin zone cuts of different Chen number are connected, whereas 0 indicates a trivial insulator. The value of $2$ can give rise to $\theta=\pi$ as it corresponds to a distribution of parity eigenvalues that amounts to twice an odd integer number $\eta$,  but we emphasize that this in general can also indicate a stacked Chern insulator \cite{Turner_2012}. It is the additional presence of the $S$ symmetry that ensures pairing and indicates that the axion possibility is realized. 
This is because the presence of $S$ symmetry ensures {\it pairing} at TRIM, eliminating the stacked Chern insulator possibility.  At $\tilde{k}=\pi$, $S$ and $I$ anti-commute and result in pairs of opposite parity eigenvalue that are of no importance for the characterization, while at $\tilde{k}=0$ these symmetries commute and thus ensure that the parity eigenvalues come as doubles of the same parity \cite{Watanabe_2018}. As a result, we observe 
that the classification gets reduced to the anticipated $\mathbf{Z}_2$ index, which is nontrivial due to the inverted pair at the $\Gamma$-point, showing that the system is a  quantized axion  ($\theta=\pi$) insulator.

We finally note that this indicator can trivially be rewritten into many other incarnations that are appearing recently. Using that the total number of occupied bands  $n=n_\mathbf{k}^++n_\mathbf{k}^-$, which are assumed to be gapped at least at the TRIM, is independent of $\mathbf{k}$ and the number of TRIMs is eight, we can express the above quantity in $n_\mathbf{k}^+$ or $n_\mathbf{k}^-$ up to a number that is an integer multiple of four. In particular, by expressing the indicator in terms of $n_\mathbf{k}^+$, we arrive at 
\begin{equation}\label{eq::indicator}
Z_4=\sum_{occ}\sum_{\Gamma_i}\frac{1+\xi_n(\Gamma_i)}{2}\qquad \text{Mod }4,
\end{equation}
where $\xi$ refers to the parity eigenvalue of the $n$th occupied state. 

We note that this characterization can generally be applied to I symmetric systems, in particular ferromagnetic systems. Crucially, however, the anti-ferromagnetic (AFM) system at hand comes with the $S$ symmetry that, as explained, collapses the classification index to $\mathbf{Z}_2$. This is in particular important for the bulk-boundary correspondence \cite{UnifiedBBc,Slager2017,Hatsugai93_Bbc,Codefects1}. Indeed, the protecting inversion is naturally broken at the edge and hence gapless edge states are not guaranteed in that case. In this sense, the AFM system is more analogous to a TRS protected TI that also has I symmetry. The latter can diagnose the $Z_2$ classification coming from TRS by the parity distribution. Surfaces that preserve TRS nonetheless have gapless states. Similarly, in the AFM axion insulator, the parties diagnose the $\theta$ term, but the $\mathbf{Z}_2$ index comes from the $S$ symmetry. As a consequence, gapped edges states only occur on $S$ breaking surfaces, see also the next Section.

Finally, to further underpin our analysis, we have also constructed the Wilson flows. As demonstrated in a series of works 
\cite{Soluyanov2011, Wi1,WindingKMZ2,bouhon2018wilson,Wi2}, invariants can generally be deduced from the winding in (nested) Wilson spectra. Heuristically, this is because they track the flow of the Wannier centers. As consistent with the previously outlined classification analysis we find a Wilson winding in the $\tilde{k}=0$ plane,
but not in the  $\tilde{k}=\pi$ plane.

\begin{figure}[!ht]
	\includegraphics[width=6 in]{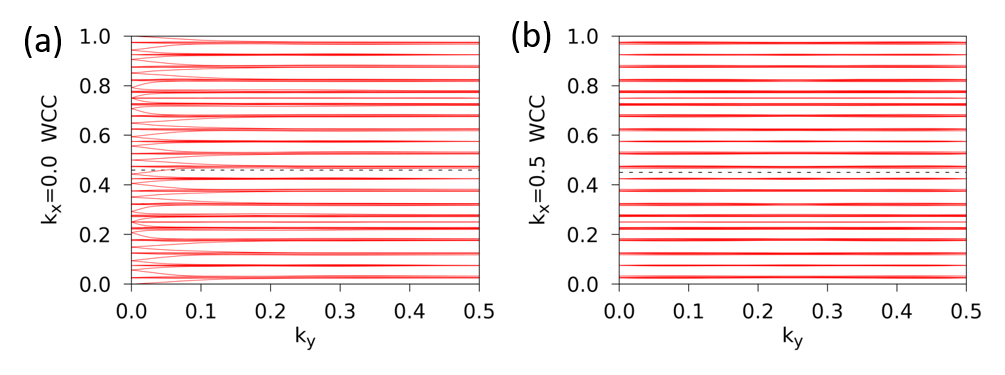}%
	\caption{\textbf{Band topology: Wannier charge centers} Evolution of Wannier charge centers (WCC) or Wilson loops for A-type AFM MnBi$_{6}$Te$_{10}$ on \textbf{a}, $k_{x}$\,=\,0.0 and \textbf{b}, $k_{x}$\,=\,0.5 plane. The 2D topological index is 1 and 0 in (a) and (b), respectively, because of odd and even number of crossing by the horizontal dashed line. The overall Fu-Kane Z2 topological index is (1;000) in the TRS protected TI context.}
	\label{fig:SI1}
\end{figure}

\subsection{DFT calculation on (110) surface}
\label{app:DFT}

\begin{figure}[!ht]
	\includegraphics[width=4 in]{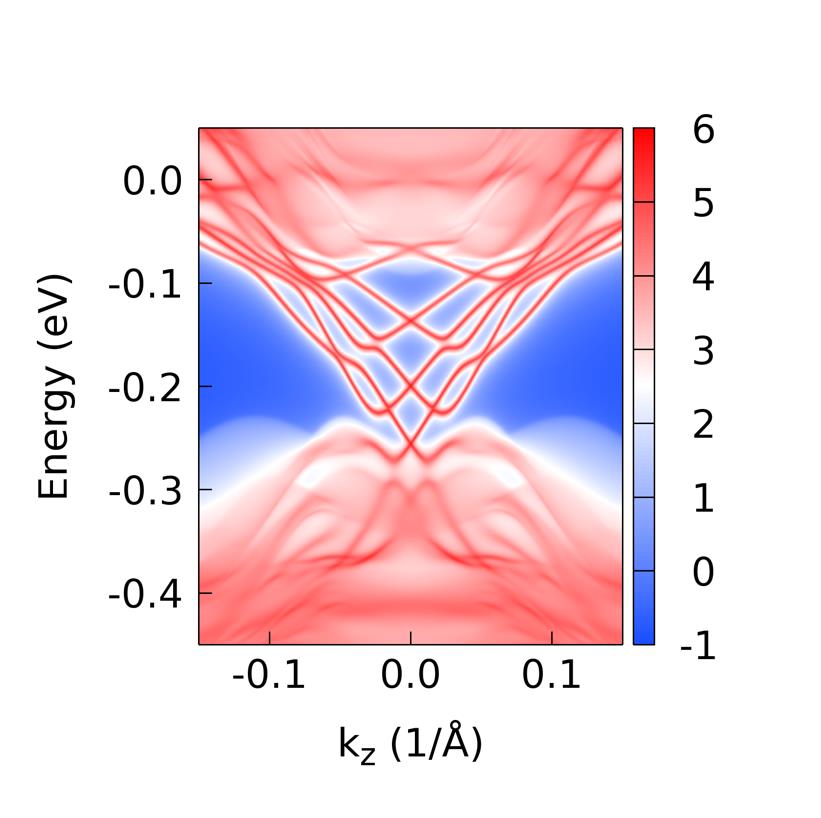}%
	\caption{\textbf{DFT calculation of $S$ preserving surface} Spectral function along $k_{z}$ direction on (110) surface of A-type AFM MnBi$_{6}$Te$_{10}$ showing the gapless surface Dirac cone for $S$ preserving surface.}
	\label{fig:SI_110}
\end{figure}

Figure\,\ref{fig:SI_110} shows a DFT calculation of A-type AFM MnBi$_{6}$Te$_{10}$ on (110) surface, where $S$ symmetry is preserved. As conistent with the previous arguments we observe a gapless surface Dirac cone, further underpinning our analysis.

\subsection{Experimental details of the ARPES measurements}

\begin{figure}[!ht]
	\includegraphics[width=6.5 in]{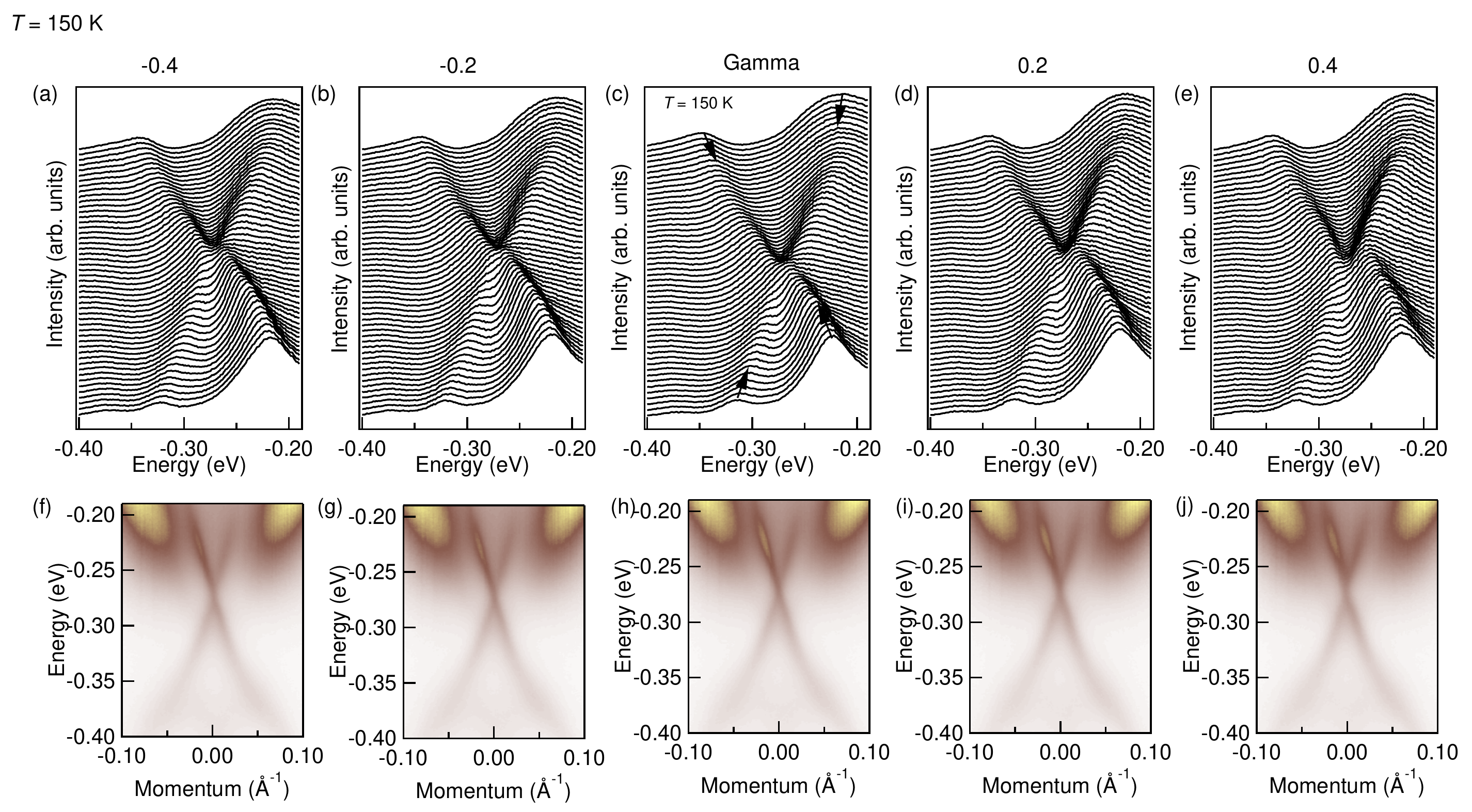}%
	\caption{\textbf{Band dispersion near $\Gamma$ point at $T$\,=\,150\,K} \textbf{a-e}, EDCs at $T$\,=\,150\,K near the $\Gamma$ point; -0.4 degree from $\Gamma$, -0.2 degree from $\Gamma$, $\Gamma$, 0.2 degree from $\Gamma$, and 0.4 degree from $\Gamma$. \textbf{f-j}, Band dispersion at $T$\,=\,150\,K near the $\Gamma$ point; -0.4 degree from $\Gamma$, -0.2 degree from $\Gamma$, $\Gamma$, 0.2 degree from $\Gamma$, and 0.4 degree from $\Gamma$.}
	\label{fig:SI2}
\end{figure}

\begin{figure}[!ht]
	\includegraphics[width=6.5 in]{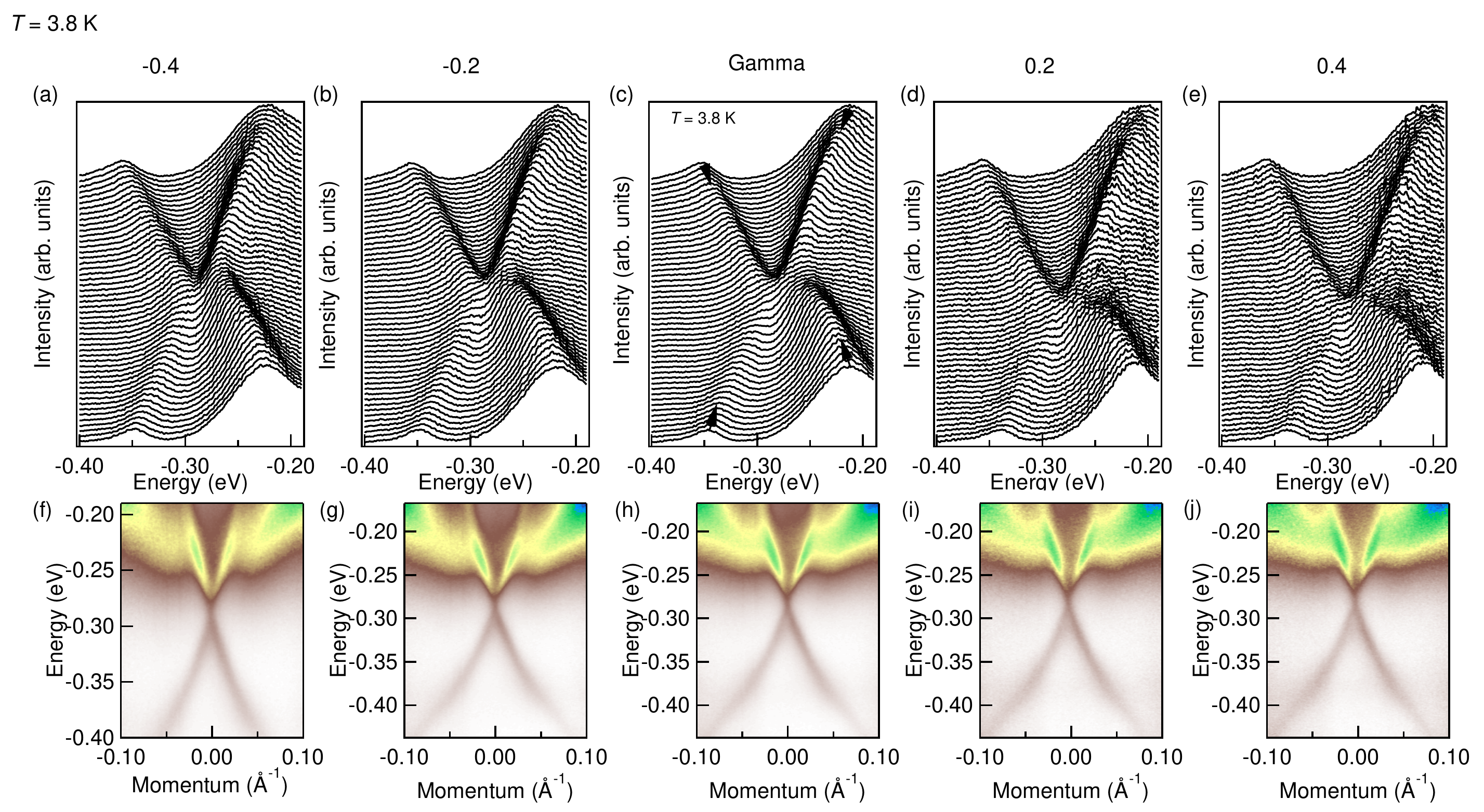}%
	\caption{\textbf{Band dispersion near $\Gamma$ point at $T$\,=\,3.8\,K} \textbf{a-e}, EDCs at $T$\,=\,3.8\,K near the $\Gamma$ point; -0.4 degree from $\Gamma$, -0.2 degree from $\Gamma$, $\Gamma$, 0.2 degree from $\Gamma$, and 0.4 degree from $\Gamma$. \textbf{f-j}, Band dispersion at $T$\,=\,3.8\,K near the $\Gamma$ point; -0.4 degree from $\Gamma$, -0.2 degree from $\Gamma$, $\Gamma$, 0.2 degree from $\Gamma$, and 0.4 degree from $\Gamma$.}
	\label{fig:SI3}
\end{figure}

We used 4 single crystals of MnBi$_{6}$Te$_{10}$ for ARPES measurements, each of them was cleaved 5 separate times. All three surface terminations were found in each of the 20 cleaves. Due to the small beam size, each termination could be quite well isolated and measured separately, although the T1Q termination seems have slight contamination with the T2Q signal, likely due to presence of small islands. T1Q termination was more rare than the other ones. We noticed slight variation of the the $E_{F}$ and the size of the low temperature gap within the TS termination. This may be due to presence of steps to adjacent terminations or small variation of sample chemistry. Samples were cleaved at $T$\,=\,3.8\,K three times to make sure the gap for the TS termination and the splitting for the T2Q termination at the low temperature is intrinsic and not due to spurious effect such as surface aging. Due to Dirac cone shape surface band, even small misalignment from $\Gamma$ will manifest itself as apparent gap. Such misalignment is relatively easy to occur in the experiment as the sample surface is not perfectly flat and small movements (e. g. due to thermal expansion then changing temperature) can change the angular orientation of surface normal. We therefore took extraordinary care to perform very fine angular scans for each measurement. Those are illustrated in Figures\,\ref{fig:SI2} and \ref{fig:SI3}. High temperature data exactly at $\Gamma$ \ref{fig:SI2}(c) shows merging of the upper and lower bands and gapless Dirac state. For even small changes of lateral momentum away from $\Gamma$ (i. e. emission angle from normal) there is clearly visible separation between upper and lower cones which may be misinterpreted as a gap. At low temperatures, Fig. \ref{fig:SI3} the upper and lower cones are separated at $\Gamma$ and this separation increases away from $\Gamma$. This is a definite proof for presence of energy gap at low temperatures. 

In order to have energy dispersions at the $\Gamma$, we performed the theta scan $\Gamma\,\pm\,\sim\,1^{o}$ with 0.2\,$^{o}$ step size for all the temperatures. Figures\,\ref{fig:SI2} and \ref{fig:SI3} show the selective theta scan for $T$\,=\,3.8\,K and 150\,K. 

\end{document}